# Effect of Pressure on Magnetism of UIrGe

Jiří Pospíšil[1,2], Jun Gouchi[3], Yoshinori Haga[1], Fuminori Honda[4], Yoshiya Uwatoko[3], Naoyuki Tateiwa[1], Shinsaku Kambe[1], Shoko Nagasaki[3], Yoshiya Homma[4] and Etsuji Yamamoto[1]

[1] *Advanced Science Research Center, Japan Atomic Energy Agency, Tokai, Ibaraki, 319-1195, Japan*

[2] *Charles University in Prague, Faculty of Mathematics and Physics, Department of Condensed Matter Physics, Ke Karlovu 5, 121 16 Prague 2, Czechia*

[3] *Institute for Solid State Physics, University of Tokyo, Kashiwa, Chiba 277-8581, Japan*

[4] *Institute for Materials Research, Tohoku University, Oarai, Ibaraki 311-1313, Japan*

## Abstract

We report the effect of hydrostatic pressure on the electronic state of the antiferromagnet UIrGe, which is isostructural and isoelectronic with the ferromagnetic superconductors UCoGe and URhGe. A series of electrical resistivity measurements in a piston-cylinder-type cell and a cubic-anvil cell were performed at hydrostatic pressures up to 15 GPa. The Néel temperature decreases with increasing pressure. We constructed a $p$-$T$ phase diagram and estimated the critical pressure $p_c$, where the antiferromagnetism vanishes, as ~ 12 GPa. The antiferromagnetic/paramagnetic transition appears to be first order. We suggest a scenario of competing antiferromagnetic inter-$J$- and ferromagnetic intra-$J^*$-chain interactions in UIrGe. A moderate increase in the effective electron mass was detected in the vicinity of $p_c$. A discussion of the electronic specific heat $\gamma$ and electron-electron correlation term $A$ using the Kadowaki–Woods relation is given.

## 1. Introduction

The strong interest in the physics of uranium intermetallics has focused on the behavior of the 5$f$ electrons on the boundary between localized and itinerant character. Among the series of materials studied extensively, ternary compounds with the chemical formula U$TX$ ($T$ : transition metal, $X$ : $p$-element) stand out. The crystal symmetry and the delicate balance between the direct interaction of the 5$f$ states and their hybridization with the electronic states of $T$ and $X$ elements lead to diverse electronic properties in these compounds. External variables such as hydrostatic pressure are an effective tool for tuning the electronic state of the uranium ion leading to superconductivity (SC), magnetism, or heavy fermion behavior.

Metamagnetic UCoAl and ferromagnetic (FM) URhAl and UCoGa crystallizing in a ZrNiAl-type structure are model examples of quantum phase transitions in FM systems. In all the cases, a tricritical points (TCPs) accompanied by first-order FM/paramagnetic transitions have been proposed[1-5] and the wing-structure phase diagrams seem to be generic for clean FM metals[6]. Of particular interest are compounds located close to the empirical Hill criterion,[7] which defines the limit of the distance of nearest uranium ions as $d_{U-U} \approx 3.5$ Å, at which 5$f$ electrons change from a localized to itinerant character. Isostructural compounds URhGe and UCoGe, which crystallize in a TiNiSi-type structure naturally fulfill this condition, and the coexistence of ferromagnetism (FM) and SC has been discovered here[8, 9]. Both FM and SC can be affected by alloying[10-14], the application of a magnetic field[15, 16], and the external pressure[17-19]. The resulting $H$-$T$-$p$-$x$ phase diagrams corroborate the close connection between FM and SC.

In this paper, we investigate UIrGe with the TiNiSi-type structure. Its importance has been highlighted by systematic study of the magnetic and structural parameters of the isostructural U$T$Ge compounds[10, 20]. It was found that the magnetism follows the Hill scenario[7] surprisingly well, see Fig. 1.

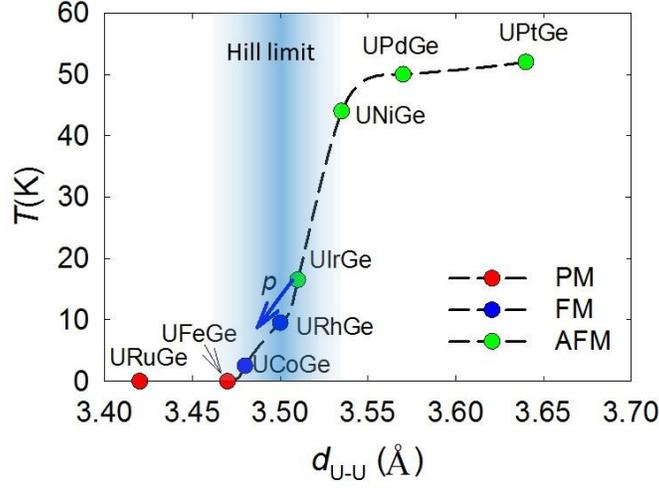

**Fig. 1.** Evolution of magnetic ordering temperature and type in orthorhombic TiNiSi-type U$T$Ge compounds as a function of $d_{U-U}$. A clear crossover from the paramagnetic (PM) to FM phase is detected in the proximity of the Hill limit. A further increase in the distance $d_{U-U}$ stabilizes the antiferromagnetic (AFM) ground state. The plot was constructed on the basis of data in Refs.10 and 20. The blue arrow tentatively indicates the expected effect of hydrostatic pressure on the magnetism of UIrGe.

UIrGe is an antiferromagnet with Néel temperature $T_N = 16.5$ K[21], unlike the isostructural and isoelectronic ferromagnets UCoGe and URhGe as shown in Fig. 1. An alloying study of the UIr$_{1-x}$Rh$_x$Ge system[22] has demonstrated that the antiferromagnetic (AFM) ground state switches to ferromagnetic (FM) upon substituting Ir by Rh. The switching mechanism of the ground state from AFM to FM has not been resolved. However, it is noteworthy that a tiny change in $d_{U-U}$ can alter the ground states of URhGe and UIrGe.

The application of a magnetic field also modifies the AFM ground state of UIrGe. The magnetic easy axis is along the $c$ axis and a sharp metamagnetic jump appears in magnetization curves at $H_{c,crit} = 14$ T followed by an induced magnetic moment above the critical field $H_{c,crit}$ of ~0.4$\mu_B$/f.u., comparable to the value of URhGe. For the $b$ axis, a field-induced transition is also observed at $H_{b,crit} = 21$ T, with even larger magnetic moment ~0.6$\mu_B$/f.u.[23].

On the other hand, the electronic heat capacity coefficient[21] of $\gamma \approx 20$ mJ/molK$^2$, which reflects the effective mass of the electrons, is considerably lower than that those of UCoGe ($\gamma = 55$ mJ/molK$^2$) and URhGe ($\gamma = 160$ mJ/molK$^2$)[20] or archetypal uranium heavy-fermion compounds[24-26]. The low $\gamma$ is considered to be connected with the opening of a large AFM gap[21].

The magnetic entropy of UIrGe $S_{mag} = 0.21R \ln2$[27] is comparable to that of URhGe[28] but significantly larger than $S_{mag} = 0.003R \ln2$ of UCoGe[9].

Figure 1 suggests that the contraction of the unit cell volume and the reduction of the distance $d_{U-U}$ by hydrostatic pressure may drive the UIrGe compound to a magnetic instability. In such a case, hydrostatic pressure should cause the closing of the AFM gap and the recovery of the high coefficient $\gamma$. Surprisingly, the $p$-$T$ phase diagram of UIrGe has not yet been established. In this paper, we report a series of high-pressure experiments in two types of pressure cells using what are considered to be UIrGe single crystals with the highest quality available. These crystals allow us to investigate the $p$-$T$ phase diagram up to 15 GPa and examine the potential development of a non-Fermi liquid (NFL) state at the critical pressure. We surmise that UIrGe may be a candidate material in which the SC state can potentially develop.

## 2. Experimental Procedure

High-quality UIrGe single crystals were grown by Czochralski pulling in a tetra-arc furnace from a polycrystalline precursor with a nominal composition of UIrGe$_{1.02}$. The single crystals were subsequently wrapped in Ta foil, sealed in a quartz tube under high vacuum, and thermally treated for 14 days at 1000°C. A precise spark erosion saw was used to cut appropriately shaped samples. All the measurements were carried out on samples in the form of a bar whose long axis was the $b$-axis. A CuBe/NiCrAl piston-cylinder-type high-pressure cell was utilized[29] to study the response of $T_N$ to hydrostatic pressure up to 1.7 GPa at temperatures down to 6 K. The cell was mounted on the cold finger of a Sumitomo SRDK-101D cryocooler. Daphne oil 7373 was used as a pressure-transmitting medium[30]. The shift of $T_{SC}$ measured on a piece of lead loaded in the cell was used as a pressure manometer[31-33]. Experiments up to 15 GPa were performed in a cubic-anvil cell. The gasket was made of semi-sintered MgO ceramic with Fluorinert as the transmitting medium inside a Teflon capsule. The pressure inside the cell was determined from the calibrated relation between the pressure at low temperatures and the applied load at room temperature. Note that the load applied at room temperature was kept constant during the cooling and heating process, which preserved the pressure in the cell throughout the experiment. Technical details are available in Ref.[34]. All the electrical resistivity measurements were performed by a conventional dc four-probe method.

## 3. Results

The AFM order of UIrGe is clearly visible as a considerable drop in the electrical resistivity at $T_N = 16.5$ K along the all three crystallographic axes-see Fig. 2(a). This contradicts the data, where a sudden increase in the electrical resistivity below $T_N$ was reported[27]. Such behavior is, however, in conflict with the electrical resistivity of polycrystalline samples in the same work.

Other specific features of the electrical resistivity of UIrGe in previous studies are a peak anomaly at $T_N$ and a shallow minimum at ~ 6.5 K[27, 35] -see Fig. 2(a). We found that both effects are sample-dependent by measuring many samples with various values of the residual resistivity ratio (*RRR*). We found that the shallow minimum at ~ 6.5 K disappears and only a smooth decrease in the electrical resistivity was observed below $T_N$ -see Fig. 2(b) in samples with *RRR* higher than 15 ($i \parallel b$) (*i*-electrical current). We did not detect features corresponding to the shallow minimum at ~ 6.5 K in the low-*RRR* samples in the heat capacity or magnetization data. The results of the heat capacity and magnetization measurements are in agreement with the published data[27, 36]. Thus, we assign the effect to an electron scattering process connected with lattice disorder or defects. High-quality samples are evidently needed for proper analyses of the features of the low temperature electrical resistivity.

The weak peak at $T_N$ in the electrical resistivity data seems to be dependent not only on the sample but also on the crystallographic direction. The effect is weak along the *b* axis, a more pronounced peak being observable along the *c* and *a* axes. We did not find any sign of a peak anomaly along the *b* axis in the high-*RRR* samples. Nevertheless, a small peak was still observable in the sample with $RRR \approx 50$ along the *c* axis. This may be evidence of the intrinsic origin of this effect. Our later analysis will show that the transformation to the AFM ordered state is connected with the opening of an AFM gap of over 60 K. A similar peak feature in the electrical resistivity data was detected for URu$_2$Si$_2$ where a complex Fermi surface reconstruction at temperature of 17.5 K at which a hidden-order transition occurs, has been discussed[37]. Because of the orthorhombic structure of UIrGe, the gap can open in specific Fermi surface regions, causing anisotropy of the size of the peak in the electrical resistivity. A detailed study of the UIrGe Fermi surface is highly desirable to clarify this effect.

A characteristic feature of U*T*Ge compounds is a value of *RRR* that is strongly sample-dependent and affected by the annealing procedure[38, 39]. This is also the case for UIrGe. The *RRR* of the as-grown single crystals is typically 2–3. Fourteen days of annealing at 1000°C increased the *RRR* parameter up to several tens but it was still strongly sample-dependent. The material with a high *RRR* could only be extracted from the single-crystal ingots by systematic measurements of a number of samples. In addition, we found that the *RRR* value is anisotropic by investigating a cubic sample (2 x 2 x 2 mm$^3$). The highest *RRR* was observed along the *b* axis. For the *a* axis, the *RRR* was about ~15% lower, while it was about ~50% lower along the *c* axis. The imprecise geometrical factor of the cubic sample does not allow a reliable estimation of the absolute value of the resistivity. However, we were able to perform the calculation along the *b* axis of various samples having the optimal shape of a bar. We obtained values of $\rho_0$ ~13 μΩcm (*RRR*-28; sample used for high-pressure experiment in cubic anvil cell), ~10 μΩcm (*RRR*-36 see Fig. 2(b) and ~5 μΩcm (*RRR*-70, will be published elsewhere) at a low temperature. The low-temperature absolute value of the UIrGe resistance is comparable with those of isoelectronic U*T*Ge systems[38, 40] and is systematically reduced by increasing the value of *RRR*.

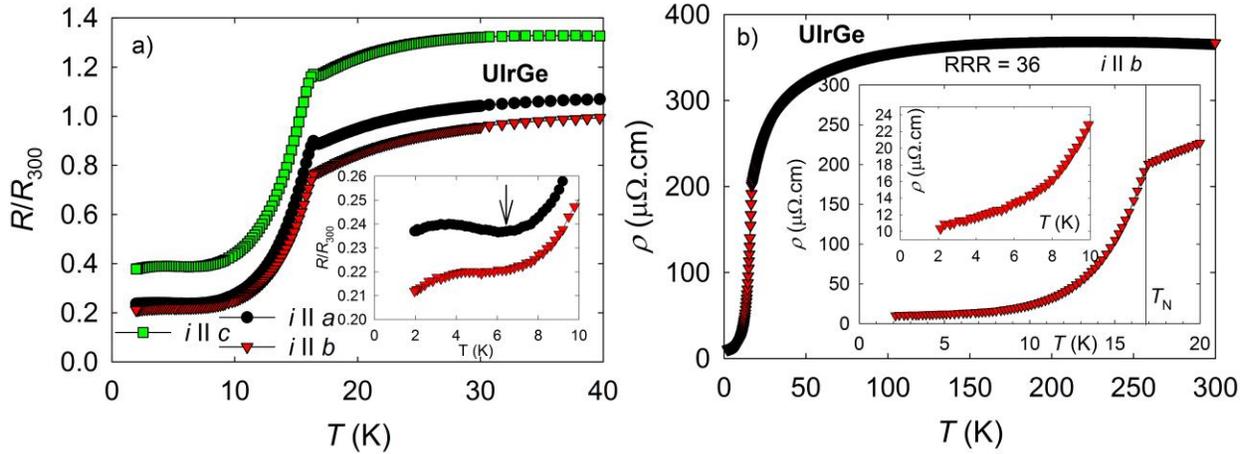

**Fig. 2.** a) Temperature-dependent electrical resistivity data of UIrGe single crystal with the current along all three directions. *RRR* along the *c* axis is reduced by about 50% in comparison with the highest *RRR* along the *b* axis. A clear peak-like anomaly appears at the $T_N$ in the resistivity when the electrical current is applied along the *a* and *c* axes. The inset shows the electrical resistivity below $T_N$ in detail. A shallow minimum is detectable around 6.5 K (black arrow). b) Electrical resistivity data of a high-quality single crystal of UIrGe in the $i \parallel b$ arrangement with $RRR = 36$. The shallow minimum at ~ 6.5 K has clearly disappeared.

A preliminary experiment in the piston-cylinder type cell at pressures up to 1.7 GPa revealed a very small shift of the ordering temperature from $T_N = 16.6$ K at ambient pressure to 16.4 K at 1.7 GPa (Fig. 3). The calculated slope of $dT_N/dp \approx -0.11$ K/GPa is small and linear extrapolation of $T_N$ indicates an anomalously high critical pressure.

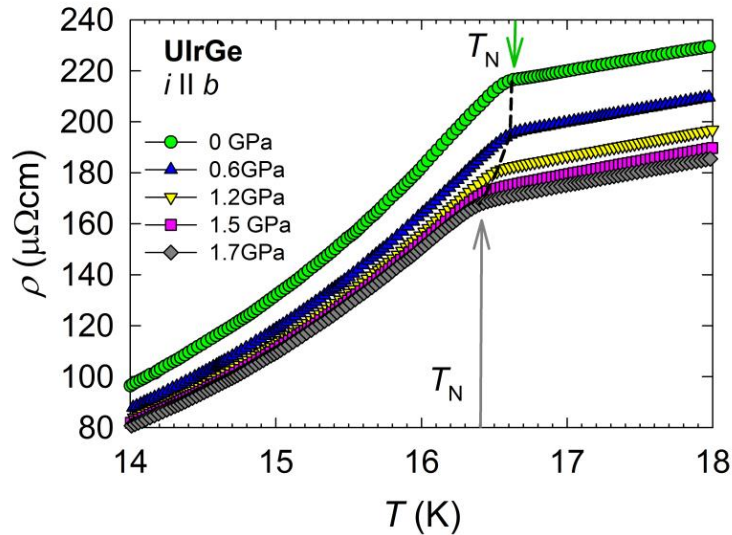

**Fig. 3.** Series of electrical resistivity isobar curves obtained by measurements in the piston-cylinder-type cell. The green and gray arrows indicate $T_N$ at ambient pressure and at 1.7 GPa, respectively. The evolution of $T_N$ is highlighted by the dashed line.

However, it is a well-known fact that the response of physical quantities to external variables is often nonmonotonic and a sudden change can appear approaching $p_c$[41, 42]. Therefore, we extended the hydrostatic pressure range up to 15 GPa using a cubic anvil cell (Fig. 4).

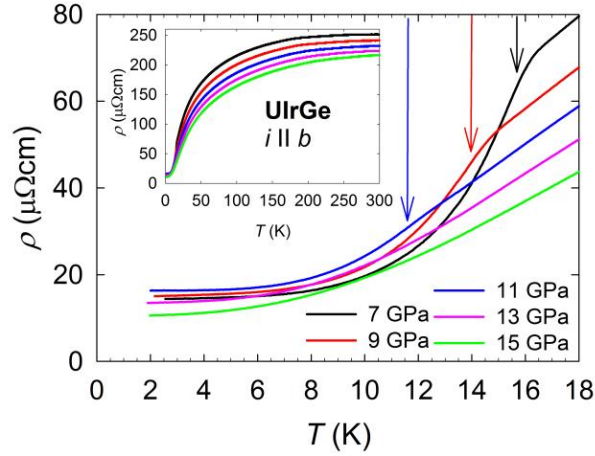

**Fig. 4.** Electrical resistivity isobar curves obtained by measurements in the cubic anvil cell up to 15 GPa. A sample with $RRR \approx 28$ was used. The inset shows the results of temperature scans up to room temperature.

The value of $T_N$ is only moderately affected up to 7 GPa. The analysis of $T_N$ up to 11 GPa is straightforward because the anomaly is detectable as a clear peak in the derivative $d\rho/dT$ at $T_N = 11.5$ K (Fig. 5). In contrast, we did not find any evidence of a magnetic transition at 13 and 15 GPa in either the resistivity or its derivative. This suggests a sudden drop in $T_N$ between 11 and 13 GPa.

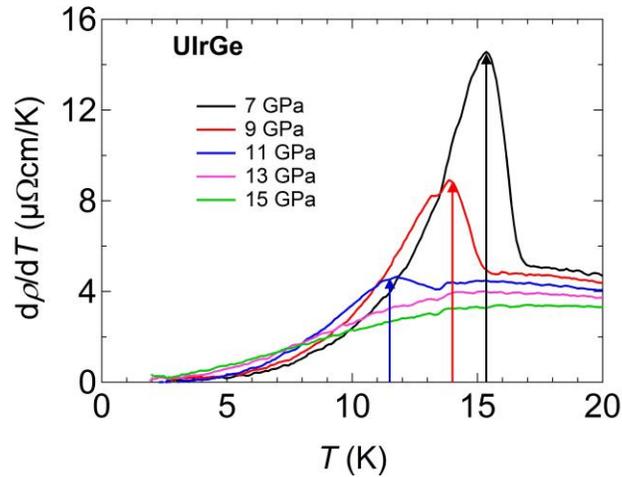

**Fig. 5.** Plot of $d\rho/dT$ data from cubic anvil cell at pressures up to 15 GPa. $T_N$ is clearly detectable at 11.5 K at 11 GPa. The arrows indicate $T_N$ at 7, 9, and 11 GPa. $T_N$ is undetectable at higher pressures.

We plotted the values of $T_N$ from all the performed measurements in the $p$-$T$ phase diagram in Fig. 6. UIrGe is the second one after UCoGe, for which it has been found that the magnetism can be suppressed by hydrostatic pressure. However, the shapes of their phase diagrams are markedly different. The steep drop to zero from a relatively high $T_N$ just above $p_c$ in UIrGe is in contrast to the gradual vanishing of the FM in UCoGe up to the intersection with the summit of the SC dome[18, 19, 43]. The sudden drop in $T_N$ at $p_c$ and the resulting rectangular $p$-$T$ phase diagram of UIrGe have the signature of a first-order transition. We surmise the scenario of a first-order transition based on single-crystal neutron diffraction. The reported AFM magnetic structure of UIrGe consists of canted FM chains along the *a* axis that are mutually antiferromagnetically coupled[44]. The FM chains along the *a* axis in UIrGe are identical to the magnetic structure of the FM SC UCoGe[45-47] and URhGe[48]. Thus, we can deduce that there are two magnetic interactions in UIrGe, antiferromagnetic inter-*J*- and FM intra-*J**-chain interactions competing with each other. The *J** has the same nature as that in FM compounds. The pressure may affect the *J*-*J** balance in UIrGe and make the FM component more important at $p_c$. However, a high-pressure neutron diffraction experiment is highly desirable to resolve the magnetic structure of UIrGe in the vicinity of $p_c$ to reveal the origin of the first-order-like transition that appeared in the recently reported $H$-$T$[40, 49] and $p$-$T$[2, 5] phase diagrams of clean FM metals.

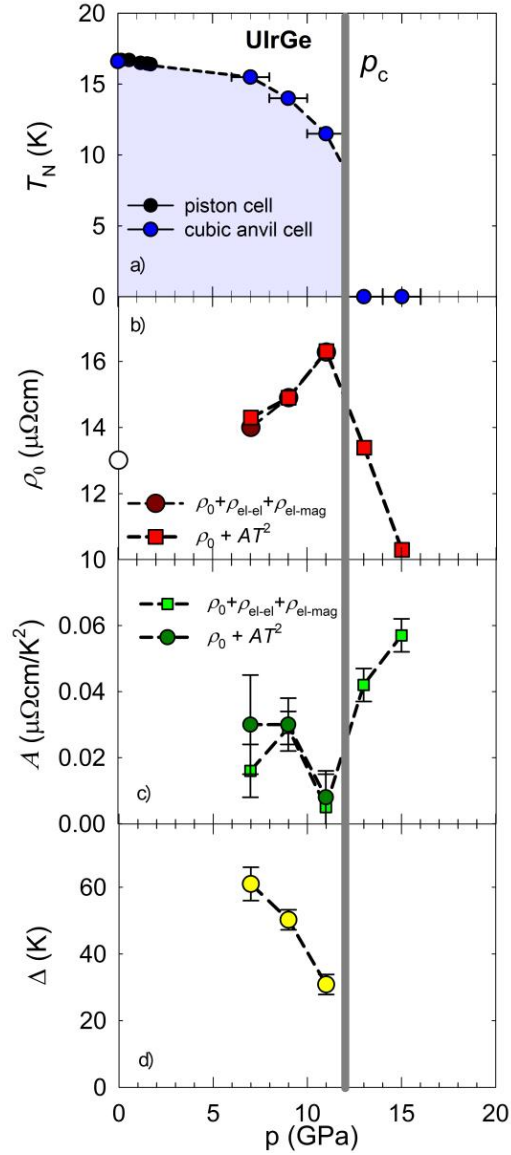

**Fig. 6.** *p-T* phase diagram of UIrGe compound. The wide gray line tentatively defines the expected critical pressure $p_c$. The sudden drop in $T_N$ at $p_c$ is followed by a maximum in $\rho_0$ (panel b) and an increase in electrical resistivity parameter *A* (panel c). $\rho_0$ and *A* below $T_N$ were evaluated using Eq.(1) and (2), and only Eq. (1) was used to fit the data above $p_c$. The $\rho_0$ at ambient pressure is displayed as an open circle in the panel b, which was obtained by evaluating *RRR* for the sample before the pressure experiment. An appropriate temperature scan of the electrical resistivity for the analysis of the other parameters is not available. Panel d shows the rapid closing of the AFM gap $\Delta$ evaluated by Eq. (2). The dashed lines are guides to the eye.

We employed the Eq. (1) to analyze the low-temperature resistivity.

$$\rho = \rho_0 + AT^2 \quad (1)$$

We investigated the residual resistivity $\rho_0$ and the electron scattering term $AT^2$. Analysis using Eq. (1) was impracticable since the electrical resistivity increases much faster than $\sim T^2$ with increasing temperature. It is reasonable to extend the eq. (1) by adding an electron-magnon scattering term[50, 51] Eq. (2) to fit the data up to $T_N$ with better agreement.

$$\rho = \rho_0 + AT^2 + BT\left(1 + \frac{2T}{\Delta}\right)\exp\left(-\frac{\Delta}{T}\right) \quad (2)$$

However, the resistivity data at 7 GPa still showed an observable deviation from the fit. This might be partly due to an anomaly around 6.5 K, which significantly affects the low temperature electrical resistivity data of the samples with low *RRR*, as mentioned above (Fig. 2). We did not observe a prominent bump in the resistivity of the sample with *RRR* = 28 used in the high-pressure experiment. However, a significant contribution might still remain and affect the analysis. The determination of *A* is therefore uncertain, as shown by the large error bars in Fig. 6. Nevertheless, the uncertainty is much smaller for the estimation of $\rho_0$. The analysis of the 7 GPa data also gave a rather large AFM gap of $\Delta \approx 60$ K.

At higher pressures, the $\rho_{el\text{-}mag}$ term becomes less important, particularly at low temperatures, so allowing us to successfully apply Eq. (1) below $\sim 4$ K. Owing the high $\theta_D$ of $\approx 180$ K[27] we assumed that the phonon contribution $\rho_{ph}$ is negligible. The results of the analysis are displayed in Figs. 7 and 8 and summarized graphically in Figs. 6(c) and (d).

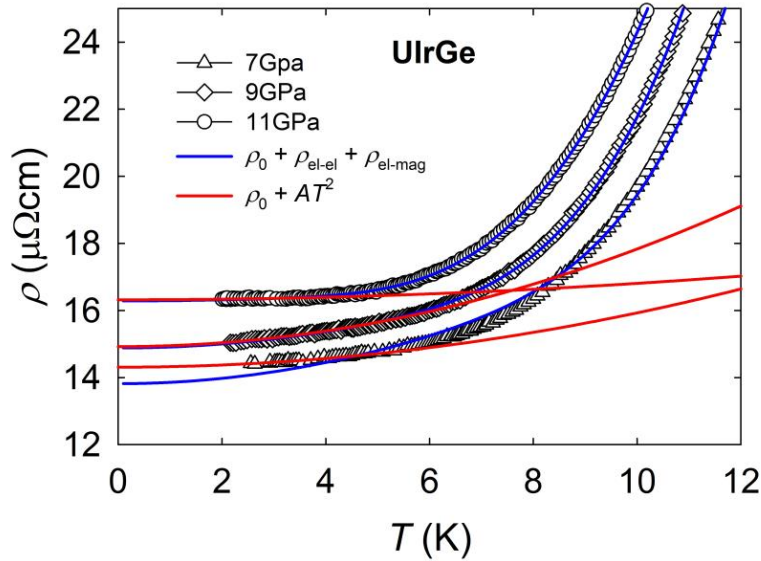

**Fig. 7.** Analysis of the electrical resistivity data at 7, 9, and 11 GPa below the expected $p_c$. Deviation of the 7 GPa data from the $\rho_0+\rho_{el\text{-}el}+\rho_{el\text{-}mag}$ fit is evident. Better agreement was found for the higher pressures of 9 and 11 GPa almost up to $T_N$. Reasonable agreement between the data and the fit by $\rho_0+\rho_{el\text{-}el}$ was found for all pressures at temperatures up to ~4 K.

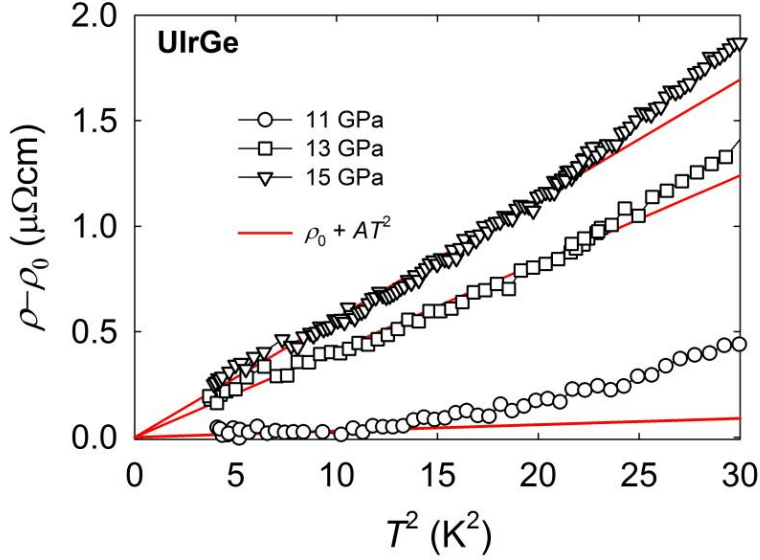

**Fig. 8.** Analysis of the low-temperature electrical resistivity data at 11, 13, and 15 GPa using Eq. (1). The plot in the $T^2$ scale shows an obvious increase in the slope of the electrical resistivity with increasing pressure, indicating the increase in the parameter $A$. Note that 13 and 15 GPa are higher pressures than the established $p_c$.

Clear evidence is found from the electrical resistivity that the parameter $\rho_0$ has a maximum at $p_c$. While $\rho_0$ is very effective for detecting various phase transformations in materials, different types of behavior have been reported. $\rho_0$ has a maximum at the critical pressure in the FM/PM transition in URhAl[2] and the recently reported $U_3P_4$[52]. In both cases a first-order FM/PM transition is induced by pressure. The $\rho_0$ for the weak itinerant FM MnSi[53] behaves similarly to that for URhAl and $U_3P_4$. On the other hand, more complex features were reported for the neighboring UCoGe[18, 43]. A local minimum appears at the critical pressure of the FM/PM transition of around 0.9 GPa. Then $\rho_0$ gradually increases to 7.2 GPa and reaches a maximum assigned to a weak valence crossover[43]. We can only speculate on the origin of the maximum $\rho_0$ in UIrGe. The enhancement of the $\rho_0$ due to the development of FM fluctuations about factor 1.4 was reported for MnSi[53] and about ~1.1 for URhAl[2]. In contrast, $\rho_0$ increases tenfold compared with that at ambient pressure in $U_3P_4$[52] due to magnetic phase separation. The $\rho_0$ for UIrGe is increased about a factor 1.25 comparable to MnSi or URhAl. Accordingly, we tentatively suppose a gradual enhancement of magnetic fluctuations in UIrGe towards $p_c$.

The development of electron-electron correlations can be deduced from the increase in the parameter $A$. The evaluation of $A$ at low pressures, especially at 7 GPa, was affected by the deviation of the fit by Eq. (2) from the data. Much better agreement was found at low temperatures using Eq. (1), especially at 11 GPa and pressures above $p_c$ (Fig. 8). The obtained value of $A \approx 0.06$ $\mu\Omega$cm/K$^2$ at 15 GPa is significantly larger than the value at 11 GPa just below $p_c$ suggesting the moderately enhanced electron effective mass in the paramagnetic state (Fig. 8). The $A$ value obtained at the highest pressure in the present experiment is still lower than that in other strongly correlated uranium systems[18, 24, 28].

We attempted to estimate the high-pressure $\gamma$ of UIrGe taking into account the Kadowaki-Woods empirical ratio[54] of $A/\gamma^2 = 1.0 \times 10^{-5}$ $\mu\Omega$cm (molK/mJ)$^2$. We obtained $\gamma_{15\text{GPa}} \approx 80$ mJ/molK$^2$.

This is four times higher than the value at ambient pressure of ~20 mJ/molK$^2$. The value of $\gamma_{15GPa}$ ≈ 80 mJ/molK$^2$ is similar to the value of $\gamma$ extrapolated from the paramagnetic limit of the heat capacity data[27]. We also attempted to calculate the ambient pressure $A$ using the reported[21] value of $\gamma \approx 20$ mJ/molK$^2$. The calculated value of $A \approx 0.004$ μΩcm/K$^2$ is in very good agreement with the value we obtained at 11 GPa just below the $p_c$. This corroborates the scenario of instantaneous closing of the large AFM gap[21] at $p_c$ and an increase in the density of 5$f$ electronic states at the Fermi level.

We consider whether hydrostatic pressure can induce the AFM to FM transformation because of the reduced volume and the decrease in of $d_{U-U}$ upon applying pressure, pushing UIrGe towards the FM region according to the Hill scenario in Fig. 1. The effect of the collapse of the AFM order followed by the development of a FM phase was recently observed in USb$_2$[42]. However, we do not have any evidence from the high-pressure electrical resistivity data that this occurs for UIrGe. This scenario contradicts the existence of a FM phase in lattice-expanded UIrGe hydride[55]. Another scenario of a FM in UIrGe can be drawn by the alloying with smaller Si atoms creating a lattice contracted paramagnet UIrSi[35, 56]. The surprising analogy between the hydrostatic and chemical pressure by Si substitution can be found in all three isoelectronic compounds. The response of UCoGe is similar to that of UIrGe[18, 19, 43]. Hydrostatic pressure increases $T_C$ only in URhGe[17]. Accordingly, the Si analogues UCoSi[56] and UIrSi[35, 56] are paramagnets. URhSi is FM[57] with a higher $T_C$ (10.5 K) than that of URhGe. It is worth noting that the anisotropy of the thermal expansion coefficients $\alpha_i$ around $T_N$ is not well known for UIrGe[58]. Knowledge of $\alpha_i$ is very useful for explaining the pressure effects on both FM and SC in UCoGe and URhGe[59, 60] with respect to the Ehrenfest relation.

## 4. Conclusions

We have successfully constructed the $p$-$T$ phase diagram of UIrGe. $T_N$ is only weakly affected by the application of hydrostatic pressure up to 7 GPa. Hydrostatic pressure has a greater effect above this value and $T_N$ seems to suddenly vanish via a first-order transition. We did not detect any sign of a magnetic transition at 13 and 15 GPa, from which we established that the critical pressure is $p_c \approx 12$ GPa. We conclude that within the group of three isoelectronic compounds, UIrGe is the second one after UCoGe, for which it has been found that the magnetism can be suppressed by hydrostatic pressure.

A high sample quality is crucial for the analysis of the low-temperature electrical resistivity data. The pressure dependence of the $A$ parameter indicates moderately enhanced electron effective mass in the paramagnetic state above $p_c$. A Kadowaki-Woods empirical analysis gave reasonable agreement between the evaluated parameters $A$ and $\gamma$ below and also above $p_c$. Both quantities change abruptly at $p_c$. This corroborates the scenario that a rather large AFM gap is closed at $p_c$ which may cause a large reconstruction of the Fermi surface and an increase of the Sommerfeld coefficient $\gamma$ from its low value at ambient pressure.

The first-order transition may support the existence of a strong FM coupling in UIrGe, whose strength increases towards $p_c$. Thus, the $p$-$T$ phase diagram of UIrGe may resemble those of clean FM metals. However, no evidence of a FM phase was detected in the present work, contrary to the prediction of the Hill plot. To definitely conclude the occurrence of a first-order transition, fine-tuning of the pressure in the vicinity of $p_c$ is necessary.

Additional high-pressure experiments down to 30 mK are in progress including the time-consuming preparation of a new batch of ultra pure UIrGe single crystals with $RRR \approx 100$ to explore the potential SC state in the vicinity of $p_c$.

## Acknowledgements

This work was supported by JSPS KAKENHI Grant Numbers 15H05884 (J-Physics), 15K05156, 15H03681, and 16K05463. The authors would like to thank Z. Fisk for fruitful discussion of the results.